\documentclass[twocolumn]{aa}
\usepackage{epsfig}
\usepackage{graphicx}
\usepackage[latin1]{inputenc}
\def\simlt{\ \raise -2.truept\hbox{\rlap{\hbox{$\sim$}}\raise5.truept   %
\hbox{$<$}\ }}
\def\simgt{\ \raise -2.truept\hbox{\rlap{\hbox{$\sim$}}\raise5.truept   %
\hbox{$>$}\ }}                                                          %

\def\be{\begin{equation}}
\def\ee{\end{equation}}
\def\newline{\hfil\break}

\def\la{\mathrel{\hbox{\rlap{\hbox{\lower4pt\hbox{$\sim$}}}\hbox{$<$}}}}
\def\ga{\mathrel{\hbox{\rlap{\hbox{\lower4pt\hbox{$\sim$}}}\hbox{$>$}}}}

\begin{document}

\title{Multi-frequency constraints on the non-thermal pressure in galaxy clusters}
   \author{S. Colafrancesco\inst{1,2}, M.S. Emritte\inst{2} , N. Mhlahlo\inst{2} and P. Marchegiani\inst{1,2}}

   \offprints{S. Colafrancesco}

\institute{
              INAF - Osservatorio Astronomico di Roma
              via Frascati 33, I-00040 Monteporzio, Italy.
              Email: sergio.colafrancesco@oa-roma.inaf.it
 \and
              School of Physics, University of the Witwatersrand,
              Johannesburg Wits 2050, South Africa.
              Email: sergio.colafrancesco@wits.ac.za
             }

\date{Received  / Accepted  }
\authorrunning {S. Colafrancesco et al.}
\titlerunning {Multi-$\nu$ constraints on cluster non-thermal pressure}

\abstract
{The origin of radio halos in galaxy clusters is still unknown and is the subject of a vibrant debate both from the observational and theoretical point of view. In particular the amount and the nature of non-thermal plasma and of the magnetic field energy density in clusters hosting radio halos is still unclear.}
{ %Aims. 
The aim of this paper is to derive an estimate of the pressure ratio $X = P_{non-th}/P_{th}$ between the non-thermal and thermal plasma in radio halo clusters that have combined radio, X-ray and SZ effect observations.}
{%Method.
From the simultaneous $P_{1.4}-L_X$ and $P_{1,4}-Y_{SZ}$ correlations for a sample of clusters observed with Planck, we derive a correlation between $Y_{SZ}$ and $L_X$ that we use to derive a value for $X$. This is possible since the Compton parameter $Y_{SZ}$ is proportional to the total plasma pressure in the cluster (that we characterize as the sum of the thermal and non-thermal pressure) while the X-ray luminosity $L_X$ is proportional only to the thermal pressure of the intracluster plasma.}
{%Results.
Our results indicate that the average (best fit) value of the pressure ratio in a self-similar cluster formation model  is  $X =0.55 \pm 0.05$  in the case of an isothermal $\beta$-model with $\beta=2/3$ and a core radius $r_c = 0.3 \cdot R_{500}$ holding on average for the cluster sample. We also show that the theoretical prediction for the $Y_{SZ}-L_X$ correlation in this  model has a slope that is steeper than the best fit value for the available data. The agreement with the data can be recovered if the pressure ratio $X$ decreases with increasing X-ray luminosity as $L_X^{-0.96}$.}
 {%Conclusions.
We conclude that the available data on radio halo clusters indicate a substantial amount of non-thermal pressure in cluster atmospheres whose value must decrease with increasing X-ray luminosity, or increasing cluster mass (temperature). This is in agreement with the idea that non-thermal pressure is related to non-thermal sources of cosmic rays that live in cluster cores and inject non-thermal plasma in the cluster atmospheres that is subsequently diluted by the ICM acquired during cluster collapse, and has relevant impact for further studies of high-energy phenomena in galaxy clusters. 
}

\keywords{Cosmology; Galaxies: clusters of galaxies; evolution; non-thermal phenomena; CMB; radio emission}

 \maketitle
%----------------------------------------------------

%%%%

\section{Introduction}

The origin of radio halos (RHs) in galaxy clusters is a long-standing but still open problem. Various scenarios have been proposed that refer to primary electron models (see, e.g., Sarazin 1999, Miniati et al. 2001), re-acceleration models (see, e.g., Brunetti et al. 2009), secondary electron models (see, e.g., Blasi \& Colafrancesco 1999, Miniati et al. 2001, Pfrommer et al. 2008),  and also geometrical projection effect models (see, e.g., Skillman et al. 2012). Each one of these models has both interesting and contradictory aspects, but each one relies on the presence of a population of relativistic electrons (and positrons) and of a large-scale magnetic field that are spatially distributed in the cluster atmosphere. 
In the following we assume that RHs are produced by an intrinsic relativistic electron population within the cluster atmosphere.
The presence of RHs in clusters requires then an additional non-thermal pressure (energy density) component in addition to the thermal pressure (energy density) provided by the intra cluster medium (ICM). 

It has been recognized that galaxy clusters hosting RHs show a correlation between their radio power measured at 1.4 GHz $P_{1.4}$ due to synchrotron emission, and their X-ray luminosity $L_X$ due to thermal bremsstrahlung emission (see, e.g.,  Colafrancesco 1999, Giovannini et al. 2000, Feretti et al. 2012) that can be fitted with a power law $P_{1.4} \propto L^{d}_X$ with slope $d$ lying in the range $1.5$ to $2.1$ (see , e.g., Brunetti et al. 2009 for a recent compilation). Such a correlation links the non-thermal particle and magnetic field energy density (pressure), related to the synchrotron radio luminosity $P_{1.4} \propto n_{e,rel} B^{(\alpha+1)/2} \nu^{-(\alpha-1)/2}  \sim P_{non-th} U_B^{(\alpha+1)/4}$ (where $\alpha$ is the slope of a power-law electron spectrum $n_{e,rel} \sim E^{-\alpha}$), with the thermal pressure $P_{th}$ of the ICM, related to the thermal bremsstrahlung X-ray emission given by $L_X \propto n_e^2 T^{1/2} \sim P_{th} t^{-1}_{cool}$, where $P_{th} \propto n_e T$ and $t_{cool} \propto T^{1/2} n^{-1}_e$.\\
An analogous correlation has been found (see, e.g., Basu 2012) between $P_{1.4}$ and the integrated Compton  parameter $Y_{SZ}$  due to the SZ effect (SZE) produced by Inverse Compton Scattering of CMB photons off the electron populations that are residing in the cluster atmosphere (see Colafrancesco et al. 2003 for details, and Colafrancesco 2012 for a recent review).
The Compton parameter $Y_{SZ} \propto \int d \ell P_{tot}$ is proportional to the total particle pressure (energy density) provided by all the electron populations in the clusters atmosphere (see Colafrancesco et al. 2003):the cluster atmosphere is thus the combination of the thermal plasma producing X-ray emission and the non-thermal plasma, at least the one producing synchrotron radio emission. Therefore, the $P_{1.4}-Y_{SZ}$ correlation links the non-thermal particle and B-field pressures, as measured by $P_{1.4}$, with the total particle pressure $P_{tot}$, as measured by $Y_{SZ}$.
For the sake of generality we write here the total particle pressure $P_{tot}$ as
\begin{equation}
P_{tot} = P_{th} + P_{non-th} = P_{th} \bigg[ 1 + X \bigg]
\label{eq.X}
\end{equation} 
where $X \equiv {P_{non-th} / P_{th}}$.\\
The correlated X-ray, SZE and radio emission from RH clusters, as shown by the $P_{1.4}-L_X$ and $P_{1.4}-Y_{SZ}$ relations, indicate that the RH cluster atmospheres must also exhibit a relation between the thermal ICM pressure $P_{th}$ and the non-thermal particle pressure  $P_{non-th}$ that can be hence constrained by observations.

In this paper we will discuss the constraints on the quantity $X$ set by the available  radio, X-ray and SZE information on a sample of RH clusters observed by Planck.
In Sect.2 we discuss the cluster data that we use in our analysis, and we discuss the theoretical approach to derive information on the pressure ratio $X$ in Sect.3. We discuss our results and draw our conclusions in the final Sect.4.

We assume throughout the paper a flat, vacuum-dominated Universe with $\Omega_m =0.32$ and $\Omega_\Lambda =0.68$ and $ H_0=67.3 km s^{-1}Mpc^{-1}$.

\section{The cluster sample}

We consider here a sample of galaxy clusters which exhibit RHs and that have also X-ray and SZE information. The cluster data that are used in our analysis are selected from the Planck Collaboration (2011) and from Brunetti et al. (2009). The cluster redshifts and the radio power $P_{1.4}$ are taken from Brunetti et al. (2009) and from Giovannini et al. (2009), the bolometric X-ray luminosity $L_X$ are taken from Reichert et al (2011) while the integrated Compton parameter $Y_{SZ}$ are taken from the Planck Collaboration (2011). We also used information on the cluster velocity dispersion collected from various authors like Wu et al. (2009), Zhang et al. (2011). As for the cluster A781 we used the information given by Cook et al. (2012) and from Geller et al. (2013)
Our final cluster sample extends the cluster sample considered by Basu (2012) by including some additional clusters for which the integrated Compton parameter is now available. 
The final RH cluster sample we use in this work is reported in Tables  \ref{tab.1} and \ref{tab.2}.
\begin{table}[width=1mm,height=60mm]
\centering
\caption{The RH clusters sample.
Notes: $^{(*)}$ No uncertainty is available. $^{(**)}$ No value is available.}
\label{tab.1}
\begin{tabular}{c c c c c c c}
\\
\hline\hline
Cluster & $z$ & $L_X $       & $ P_{1.4} $         \\
          &      & $(10^{44} erg s^{-1})$ & $ (10^{24} W/Hz)$\\
\hline
1ES0657 & 0.2994 &$ 65.2\pm0.90$ &$ 28.21\pm1.97$ \\
RXCJ2003 & 0.3171 & $27.23\pm4.95 $ & $12.30\pm0.71$ \\
A2744 & 0.3080 &$ 22.12\pm1.70$ & $17.16\pm1.71$\\
A2163 & 0.2030 & $64.1\pm5.3$ & $18.44\pm0.24$\\
A1300&0.3071&$18.0\pm1.50$&$6.09\pm0.61$\\
A0665&0.1816&$21.7\pm2.00$&$3.98\pm0.39$\\
A773&0.2170&$20.9\pm1.60$&$1.73\pm0.17$\\
A2256&0.0581&$10.7\pm0.90$&$0.68\pm0.12$\\
Coma&0.0231&$10.44\pm0.28$&$0.72\pm0.06$\\
A0520&0.2010&$20.1\pm0.70$&$3.91\pm0.39$\\
A209&0.2060&$13.3\pm1.10$&$1.19\pm0.26$\\
A754&0.0535&$12.94\pm0.99$&$1.08\pm0.06$\\
A401&0.0737&$16.8\pm1.0$&$0.22^{(*)}$\\
A697&0.282&$41.9\pm2.3$&$1.91^{(*)}$\\
A781&0.3004&$6.3\pm1.0$&$4.07^{(*)}$\\
A1995&0.3186&$17.1\pm0.2$&$1.35^{(*)}$\\
A2034&0.113&$9.5\pm1.0$&$4.37^{(*)}$\\
A2218&0.1756&$11.1\pm0.8$&$0.40^{(*)}$\\
A1689&0.1832&$28.4\pm1.0$& $-^{(**)}$\\
MACSJ0717&0.5548&$84.18\pm1.01$&$50.0\pm10$\\
A1914 & 0.1712 & $21.70\pm1.1$ & $5.24\pm0.24$\\
A2219 & 0.2256 & $45.10\pm2.3$ & $1.23\pm0.57$\\
A2255 & 0.0806 & $6.50\pm0.7$ & $0.89\pm0.04$\\
\hline
\end{tabular}
\end{table}
Table \ref{tab.1} reports the values of the cluster radio halo power $P_{1.4}$, the bolometric X-ray luminosity $L_X$ and the redshift $z$. Table \ref{tab.2} reports for the same clusters in Tab. \ref{tab.1} the values of the integrated Compton parameter $Y_{SZ}$.
\begin{table}[width=1mm,height=60mm]
\centering
\caption{Cluster values for $Y_{SZ}$ and angular size $\Theta_X$ as given by the Planck Collaboration (2011).}
\label{tab.2}
\begin{tabular}{c c c}
\\
\hline\hline
Cluster & $Y_{SZ}$  & $\Delta Y_{SZ}$ \\
         & $(arcmin^{2})$ & $ (arcmin^{2})$\\
\hline
1ES0657 &0.0067 &0.0003\\
RXCJ2003 &0.0027 &0.0004\\
A2744 & 0.0042&0.0005\\
A2163 & 0.0173&0.0007\\
A1300 & 0.0035&0.0005\\
A0665 & 0.006&0.0005\\
A773 & 0.0038&0.0004\\
A2256 & 0.0242&0.0009\\
Coma & 0.1173&0.0054\\
A0520 & 0.0046&0.0006\\
A209 & 0.0053&0.0005\\
A754 & 0.033&0.0012\\
A401 & 0.0193&0.0016\\
A697 & 0.0051&0.0005\\
A781 & 0.0017&0.0003\\
A1995 & 0.0015&0.0003\\
A2034 & 0.0055&0.0008\\
A2218 & 0.0044&0.0003\\
A1689 & 0.0071&0.0008\\
MACSJ0717 & 0.0028&0.0004\\
A1914 & 0.0057&0.0005\\
A2219 & 0.0085&0.0005\\
A2255 & 0.0103&0.0006\\
\hline
\end{tabular}
\end{table}
%
%{\bf \section{ New Cosmological values}
Since the values of the cosmological parameters have been updated to the new values given by the Planck Collaboration (2013), we re-scale the $L_X$ and $P_{1.4}$ values in order to accommodate them to the new cosmological model used here. We rescale our $P_{1.4}$ as follows
\begin{equation}
\frac{P_{1.4}'}{P_{1.4}} = \frac{D_L'^2}{D_L^2}
\end{equation}
and for the bolometric luminosity $L_X$ we obtain
\begin{equation}
\frac{L_{X}'}{L_{X}} = \frac{D_L'^2}{D_L^2}
\end{equation}
where the dashes represent the new value. 
As for the Compton parameter we just re-calculated ${Y_{SZ} D_{A}^2 }$ using the new cosmological values.

\subsection{Correlations}

\begin{figure}
\centering
\includegraphics[width=80mm,height=90mm]{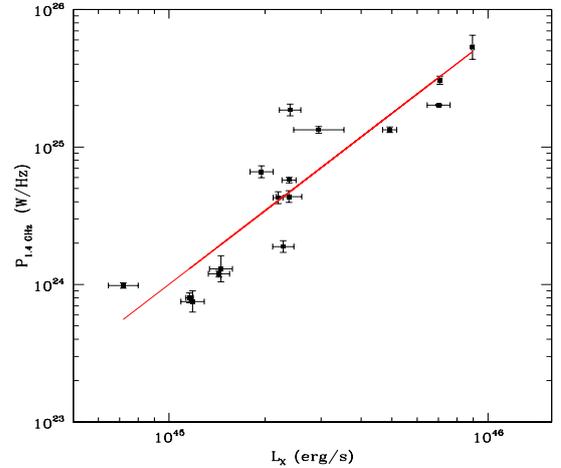}
\caption{The best fit power-law relation $P_{1.4} = C \cdot L^d_X$ for our cluster sample. The best-fit parameters are $ d=1.78 \pm 0.07$ and $Log\ C=-56.04 \pm 3.18$.}
\label{fig.1}
\end{figure}

\begin{figure}
\centering
\includegraphics[width=80mm,height=90mm]{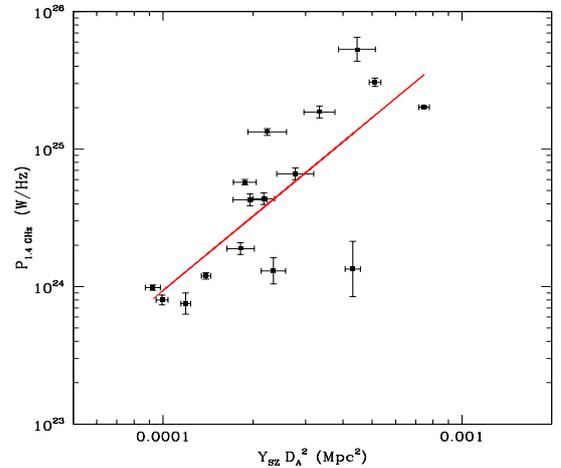}
\caption{The best fit power-law relation $P_{1.4} = B (Y_{SZ} D_A^2)^a$ for our cluster sample. The best-fit parameters are  
$a=1.80\pm 0.10$ and $ Log B= 31.16\pm 0.36$.}
\label{fig.2}
\end{figure}

The sample of RH clusters we consider in this paper exhibits the $P_{1.4}-L_X$ and $P_{1.4}-Y_{SZ}$ correlations shown in Figs. \ref{fig.1} and \ref{fig.2}.
%We show in Figs.\ref{fig.1} and \ref{fig.2} the correlations of cluster data in the planes $P_{1.4}-L_X$ and $P_{1.4}-Y_{SZ}$
Because of the common variable $P_{1.4}$ in both correlations shown in Figs. \ref{fig.1} and \ref{fig.2}, a correlation between $Y_{SZ}$ and $L_X$ is then expected theoretically and it is actually found in the data (see Fig.\ref{fig.3}).
%

%%% New version
In order to fit the $P_{1.4}$  -- $ L_{X}$, the $P_{1.4}$  -- $ Y_{SZ}$ and the $Y_{SZ}$ -- $ L_x $ correlations, we have adopted the approach of Akritas and Berchady (1996). According to this approach, in order to fit a straight line $y=m x + c$ to a data set, the slope and the intercept are given as follows
\begin{equation}
\displaystyle{m=\frac{\sum_{i=1}^{N} (x_i - \bar{x})(y_i - \bar{y})-\sum_{i=1}^{N} \sigma_{y,i} \sigma_{x,i}}{\sum_{i=1}^{N} ( x_i - \bar{x} )^2 - \sum_{i=1}^{N} \sigma_{x,i}^2}}
\end{equation}
and
\begin{equation}
c=\bar{y}-m\bar{x}
\end{equation}
where $\bar{x}$ is the mean of $x$ and same for $y$. $ \sigma_{x,i}$ and $\sigma_{y,i}$ are the errors in $x$ and $y$.
A proper treatment of the error propagation shows that the variance in the slope and in the normalization of the best-fit line can be computed as
\begin{equation}
\sigma_m^2=\sum_{j=1}^{N} \Bigg( \frac{1}{W(y_j)} \Bigg(\frac{\partial m}{\partial y_j} \Bigg)^2 +\frac{1}{W(x_j)} \Bigg(\frac{\partial m}{\partial x_j}\Bigg)^2 \Bigg)
\end{equation}
\begin{equation}
\sigma_c^2=\sum_{j=1}^{N} \Bigg( \frac{1}{W(y_j)} \Bigg(\frac{\partial c}{\partial y_j} \Bigg)^2 +\frac{1}{W(x_j)} \Bigg(\frac{\partial c}{\partial x_j}\Bigg)^2 \Bigg) \;.
\end{equation}
where
\begin{equation}
W (x_i) =\frac{1}{ \sigma_{x,i}^2}
\end{equation}
and 
\begin{equation}
W (y_i) =\frac{1}{ \sigma_{y,i}^2}
\end{equation}
In addition to the previous analysis of the variance in the slope and of the normalization, a further treatment is needed here to take into account the intrinsic scatter in the data. 
In order to estimate this intrinsic scatter we follow the method outline in Akritas and Berchady (1996) which summarize as follows
\begin{equation}
R_{i}=y_i-c-m x_i
\end{equation}
where $R_{i}$ is the residual. Then the intrinsic scatter $ \sigma_{0}^2$ is estimated as follows
\begin{equation}
\sigma_{0}^2 =\frac{ \sum_{i=1}^{N} ( R_i - \bar{R} )^2 - \sum_{i=1}^{N} \sigma_{y,i}^2}{N-2}
\end{equation}
The $\chi^2$ is then written as
\begin{equation}
\chi^2=\displaystyle{\sum_{i=1}^{N} \frac{(y_i-m x_i-c)^2}{\sigma_{y_i}^2+m^2 \sigma_{x_i}^2+\sigma_0^2}}
\end{equation}
where 
$\sigma^2_{x_i}$ and $\sigma^2_{y_i}$ are the corresponding variances of $x_i$ and $y_i$, respectively.\\
Our analysis yields the correlations $P_{1.4} = C \cdot L_X^d$ with best-fit parameters $Log\ C=-56.04 \pm 3.18$ and $d=1.78 \pm 0.07$, and  $P_{1.4} = B \cdot (Y_{SZ} D^2_A)^a$ with best-fit parameters $Log\ B=31.16 \pm 0.36$ and $a=1.80 \pm 0.10$. 
The results obtained here are quite consistent with those obtained by Brunetti et al. (2009), where $d$ was found to be in the range of $ 1.5 \div 2.1 $ and $Log\ C$ in the range $-55.4 \div -60.85$, and with the analysis of  Basu (2012), who obtained $Log\ B=32.1 \pm 1$ and $a=2.03 \pm 0.28$ for the Brunetti et al. (2009) RH sample.\\
The same data also exhibit a correlation between the Compton parameter $Y_{SZ} D^2_A$ and the X-ray bolometric luminosity $L_X$. Our analysis of this power-law correlation $Y_{SZ} D^2_A = c L^m_X$ provides best fit slope of $m=0.89\pm0.05$ and a normalization of $Log c=-44.11\pm2.23$.

\section{Theoretical analysis}

The characteristic quantities that describe the galaxy cluster structure are defined in a simple self-similar model (see, e.g., Arnaud et al. 2010, and references therein).
We first derive a relation between the Compton parameter $Y_{sph,500}$ and the bolometric X-ray luminosity $L_X$ for a general cluster in the case of a constant ICM density over $R_{500}$. Then, following the same approach, we derive the same relation for the more realistic case of an isothermal $\beta$-model for the radial profile of the ICM number density. The final results presented in this paper refer to the case of the isothermal $\beta$-model.

The mass $M_{500}$ is defined as the mass within the radius $R_{500}$ at which the
mean mass density of the cluster is $500$ times the critical density, $\rho_c (z)$, of the universe at the cluster redshift
\begin{equation}
M_{500} = {4 \over 3} R^3_{500} \cdot 500 \rho_c(z)
\label{eq.m500}
\end{equation}
with $\rho_c (z) = 3H^2(z)/(8 \pi G)$. Here $H(z)$ is the Hubble constant given by
$H(z) = H(0) [\Omega_M (1 + z)^3 + \Omega_{\Lambda} ]^{1/2}$ and $G$ is the Newtonian constant of gravitation. 
%We compare the radius $R_{500}$ defined in eq.(\ref{eq.m500}) with the value derived from Planck SZE at a (Planck Collaboration 2011) using the angular size %$\Theta_X$ for our cluster sample reported in Table \ref{tab.2}. 
%The theoretical values of $R_{500}$ for our cluster sample yield quite consistent result with the Planck data within a maximum of $20 \%$  difference. 

%We assume a value of $n_{e,500}= 1.1 \times 10^{-3} cm^3$ for this uniform density over $R_{500}$. In fact the density has to be chosen so %as not to have $X$ negative.
%\subsection{Number density $n_{e500}$}
%
%The electron number density is defined as
%\begin{equation}
%n_{e,500}=\frac{\rho_{g,500}}{\mu_e m_p} 
%\end{equation}
%(see Arnaud et al. 2009), where $\frac{\rho_{g,500}}{\mu_e m_p} =500 f_B \rho_c (z)$ with $f_B=0.175$  being the baryonic-fraction of the %universe and $\rho_c (z)$ is the critical density of the universe at the corresponding redshift.

%\subsection{The Temperature $ T_X$}
%
%We define the ICM temperature of a cluster as
%\begin{equation}
%K_B T=\frac{G\mu m_p M_{500}}{2 R_{500}}
%\end{equation}
%where
%\begin{equation}
%M_{500}=\frac{4}{3} \pi R_{500}^3 500 \rho_c (z)
%\end{equation}
%is the total mass enclosed in the radius of $R_{500}$ at which the mean density is $500$ times the critical density of the %universe at that redshift (Arnaud et al 2009)).

The characteristic thermal pressure of the cluster ICM at $R_{500}$ is defined as
\begin{equation}
P_{500} = n_{e,500} kT_{500}
\label{eq.p500}
\end{equation}
where $n_{e,500}$ and $kT_{500}$ are the thermal ICM electron numbers density and temperature, respectively.
The electron number density is defined as
\begin{equation}
n_{e,500}=\frac{\rho_{g,500}}{\mu_e m_p} 
\end{equation}
(see Arnaud et al. 2010), where $\rho_{g,500} = 500 f_B \rho_c (z)$ with $f_B=0.175$  being the baryonic-fraction of the universe, $m_p$ is the proton mass and $\mu_e \approx 1.14$ is the mean molecular weight of the gas per free electron. 
The temperature $T_{500}$ writes as
\begin{equation}
kT_{500} = {\mu m_p G M_{500}  \over 2 R_{500} }
\label{eq.t500}
\end{equation}
where $\mu$ is the mean molecular weight. The temperature $T_{500}$ is then the uniform temperature  of an isothermal sphere with mass $M_{500}$ and radius $R_{500}$.

The characteristic bremsstrahlung X-ray luminosity (see, e.g. Rybicki and Lightman 1985) of a cluster can be written as
\begin{equation}
L_{X,500} = C_2 {4 \pi \over 3} R^3_{500} n^2_{e,500} T^{1/2}_{500} \propto R^3_{500} P_{500} t^{-1}_{cool}
\label{eq.lx}
\end{equation}
where $t_{cool}= T^{1/2}_{500} / n_{e,500}$.
The normalization constant $C_{2}$ in the previous eq.(\ref{eq.lx}) takes the value of $1.728 \times 10^{-40} W s^{-1} K^{-\frac{1}{2}}  m^{3}$.
%where ${\bar g}$ is the average Gaunt factor for the thermal bremsstrahlung emissivity (see, e.g. Rybicki and Lightman 1985).

%\subsection{ X-ray Luminosity ${ L_X}$}
%
%The X-ray bremsstrahlung luminosity of each cluster is defined, assuming the the hot ICM is fully ionized, as follows 
%\begin{equation}
%\large
%L_x=C_2  T^\frac{1}{2} n_e^2 \frac{4}{3} \pi R_{500}^3
%\end{equation}
%where $T$ is the ICM temperature and $n_e$ is its electron number density.
%If we assume, for simplicity, a constant electron number density up to $ R_{500}$, the normalization constant $C_{2}$ takes %the value of $1.44 \times 10^{-40} W s^{-1} K^{-\frac{1}{2}} m^{3}$.

The characteristic integrated spherical Compton parameter calculated within the radius $R_{500}$ can be written as
\begin{equation}
Y_{sph,R500}  =  { \sigma_T \over m_e c^2}  {4 \pi \over 3}  R^3_{500} P_{500} (1+X) 
\label{eq.ysz}
\end{equation}
We have previously denoted with $Y_{SZ} D_A^2$ the cylindrical Compton parameter within a radius $5 \cdot R_{500}$ and here we introduce as $Y_{sph,R500}$ the spherical Compton parameter within the radius $R_{500}$. We note that the spherical Compton parameter is equal to the cylindrical Compton parameter within the radius $5 \cdot R_{500}$ as pointed out by Arnaud et al. (2010).
%The integrated Compton parameters $Y_{500}$ writs as
%\begin{equation}
%Y_{500}=\frac {\sigma}{m_e c^2} \frac{4}{3}\pi R_{500} ^3 n_{e500} k_B T \; .
%\end{equation}
Since the Integrated Compton parameter given for the SZE data by the Planck Collaboration (2011) is measured at a radius of $5 \cdot R_{500}$ , we scale the data of the Compton parameter given by the Planck Collaboration (2011) down to $R_{500}$ using the relation given in Arnaud et al. (2010) as follows
\begin{equation}
Y_{sph,R500} = {I(1) \over I(5)} Y_{sph,5R500}
\end{equation}
where the value of $I(1)=0.6552$ and $I(5)=1.1885$. These values are given in the Appendix of Arnaud et al. (2010).

\subsection{The $Y_{SZ}-L_X$ relation}

\begin{figure}
\centering
\includegraphics[width=80mm,height=90mm]{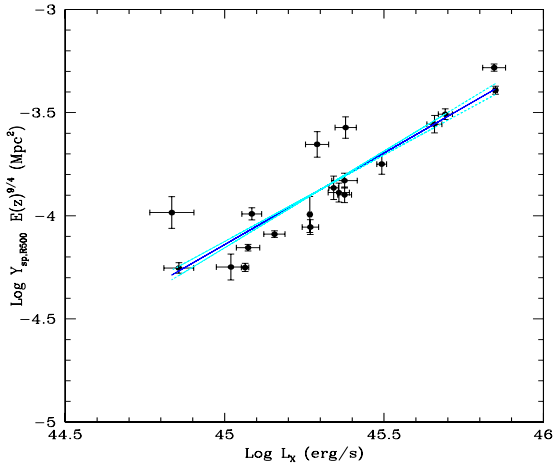}
\caption{The correlation between $Y_{sph,R500} \cdot E^{9/4}(z)$ and $L_X$ correlation for the cluster sample we consider in this paper. The best-fit power-law relation has a slope $m=0.89\pm0.05$ and a normalization of $Log c=-44.11\pm2.23$ and it is shown by the blue solid curve. The dashed curves showing the uncertainties in the slope of the correlation are shown in cyan.}
\label{fig.3}
\end{figure}

We derive here the correlation between the spherical integrated Compton parameter $Y_{sph,R500}$ and the bremsstrahlung bolometric X-ray Luminosity $L_X$ by using the simple self similar cluster model previously discussed. The correlation between the spherical Compton parameter $Y_{sph,R500}$ and the bolometric X-ray luminosity $L_X$  shown by our cluster sample is given in Fig.\ref{fig.3}. 
We derive the theoretical relation between the integrated Compton parameter and the bolometric X-ray luminosity by using eqs. (\ref{eq.lx})-(\ref{eq.ysz}) first assuming a distribution of the plasma with a constant density over $R_{500}$  and then we generalize the same relation to a more realistic density distribution. The slope of the correlation is in fact independent of the assumed cluster density profile while it affects only its  normalization.
Combining eqs. \ref{eq.m500}-(\ref{eq.ysz}) and eliminating $R_{500}$,  we obtain
\begin{eqnarray}
Y_{sph,R500} E^\frac{9}{4}(z)= & \displaystyle{\frac{\displaystyle{\Bigg(1+X \Bigg) \frac{8 \pi^2}{9}}\displaystyle{\Bigg(\frac{\sigma_T G m_p \mu 500 \rho_c n_e}{m_e c^2}\Bigg)}}{\Bigg[\displaystyle{\frac{4 \pi}{3}C_2 n_e^2}\displaystyle{\Bigg(\frac{2}{3 k_B} \pi G \mu 500 \rho_c m_p \Bigg)^\frac{1}{2}}\Bigg ]^\frac{5}{4}}} \times \nonumber \\
      & \times \displaystyle{\Bigg(\frac{L_X}{10^7 erg/s}\Bigg)^\frac{5}{4}}
\label{eq.ysz_lx_nconst}
\end{eqnarray}
The quantity $E(z)$ is the ratio of the Hubble constant at redshift
$z$ to its present value, $H_0$, i.e., $E(z) = [\Omega_M (1+z)^3 + \Omega_{\Lambda}]^{1/2}$. 
%Several theoretical arguments (refs...) indicate values of non-thermal to thermal pressure ratio $X$ in RH clusters in the range $0.1 \div 1$ (see, e.g., %Colafrancesco et al. 2003; ...), even though there could be large variations from cluster to cluster due to their specific non-thermal activity and relativitic %particle content. 
In order to estimate the best-fit value of $X$ from our cluster sample we minimize the $\chi^2$ for the $Y_{SZ} - L_X$ relation with respect to the value $X$.
We first consider the case in which the pressure ratio $X$ is constant and therefore, the non-thermal pressure has the same radial distribution of the thermal plasma in the cluster. We will discuss in Sect. 4 below the impact of this assumption on our results.

%\subsection{ Isothermal beta Model }
We now compute the same correlation $Y_{sph,R500} - L_X$ by using a more realistic, but still simple,  isothermal  $\beta$-model (see, e.g., Sarazin 1988 for a review) in which the ICM is assumed to be in hydrostatic equilibrium with the pressure balancing gravity. 
Following Ota and Mitsuda (2004), the equation for hydrostatic equilibrium writes as
\begin{equation}
\frac{k_{B}T}{\mu m_p} \Bigg(\frac{d\ ln\ \rho_{gas}}{d\ ln\ r} \ + \ \frac{d\ ln\ T}{d\ ln \ r} \Bigg)\ =\ - \frac{G M(r)}{r}
\label{eq.hydroeq}
\end{equation}
where $M(r)$ is the total mass enclosed in a radius $r$. In a simple $\beta$-model density profile $ \rho_{g}(r)=\rho_{g,0} \bigg[1+ \bigg(\frac{r}{r_{c}}\bigg)^2 \bigg]^{-\frac{3\beta}{2}}$ where $\rho_{g,0} $ is the central gas density, $r_c$ the core radius and $\beta$ takes usually values $\sim 0.5 \div 1$, the mean total  density, $\bar{\rho} (r)$ inside a radius of $r$ is given by
\begin{equation}
\bar{\rho}(r)=\frac{3 M(r)}{4 \pi r^3}=\frac{\rho_{0}}{1+({\frac{r}{r_c}})^2}
\end{equation}
where $\displaystyle{\rho_{0}=\frac{9 k_B T \beta}{4 \pi G \mu m_p {r_c}^2}}$  is the central total density of the cluster. 
From this one can write the central gas number density as
\begin{equation}
n_{e0,g}=\frac{f_B \rho_{0}}{\mu_{e} m_p}
\end{equation}
Then using Eqs. (17) and (18) and writing $r_c=\lambda R_{500}$ one can cast the central gas number density as
\begin{equation}
n_{e0,g}=\frac{3\beta f_{B}500 \rho_{c}}{2 \lambda^2 \mu_{e} m_p}
\end{equation}
Several values of $\lambda$ have been used by different authors (see, e.g., Bahcall 1975, Sarazin 1988, Dressler 1978) suggesting that for typical rich clusters the value of $\lambda$ is in the range $0.1 \div 0.25$. For X-ray clusters the value of $\lambda$  can even go up to $0.3$. We adopt here the value of $\lambda=0.3$ which gives consistent values of $X \geq 0$ for the majority of the RH clusters in our sample. We notice, in fact, that the value of $X$ is sensitive to the central number density in the formalism we adopt here, with large values of the central density leading to large values of $X$. We stress that this description assumes that the non-thermal plasma is distributed spatially as the thermal ICM, and that the pressure ratio $X$ is therefore spatially constant. Relaxing this assumption can provide slightly different results that we will discuss in a further analysis of the radial distribution of the cluster pressure structure (Colafrancesco et al. in preparation).

Under the $\beta$-model density profile assumption, the spherical integrated Compton parameter and the X-ray luminosity within $R_{500}$ can be written as
\begin{eqnarray}
Y_{sph,R500} E(z)^{-4} & = &\big(1+X\big)\frac{8\pi^2}{3} \frac{\sigma_T}{m_e c^2}  \nonumber \\
                                          &     & \times G \mu m_p 500\rho_c n_{e0,g} \lambda^3  R_{500}^5 V_1(\lambda)
\end{eqnarray}
and
\begin{eqnarray}
L_X E(z)^{-5} & = & 4\pi C_2 \bigg(\frac{2\pi}{3 k_B} G \mu m_p 500 \rho_c \bigg)^{\frac{1}{2}}  \nonumber \\ 
                      &     & \times n_{e0,g}^2 \lambda^3 R_{500}^4 W_1(\lambda)
\end{eqnarray}
where
\begin{equation}
 \displaystyle{V_1(\lambda)=\int_{0}^{\frac{1}{\lambda}} \bigg(1+u^2 \bigg)^{-\frac{3\beta}{2}} u^2 du}
\end{equation}
and
\begin{equation}
\displaystyle{W_1(\lambda)=\int_0^{\frac{1}{\lambda}} \bigg(1+u^2\bigg)^{-3\beta} u^2 du} \;.
\end{equation}

% Additional formulae
In order to clarify the main dependence of the integrated Compton parameter from the bolometric X-ray luminosity we write eqs.(25-26) into a compact form  similar to eq.(20)
\begin{equation}
Y_{sph,R500} E(z)^{9/4}  = \bigg[ {(1+X) Y_0 \over L^{5/4}_0}\bigg] L_X^{5/4}
\end{equation}
where we have defined the following quantities
\begin{equation}
Y_0 = \frac{8\pi^2}{3} \frac{\sigma_T}{m_e c^2} G \mu m_p 500\rho_c n_{e0,g} \lambda^3  V_1(\lambda)
\end{equation}
\begin{equation}
L_0 = 4\pi C_2 \bigg(\frac{2\pi}{3 k_B} G \mu m_p 500 \rho_c \bigg)^{\frac{1}{2}}  n_{e0,g}^2 \lambda^3  W_1(\lambda)
\end{equation}

\begin{figure}
\centering
\includegraphics[width=80mm,height=90mm]{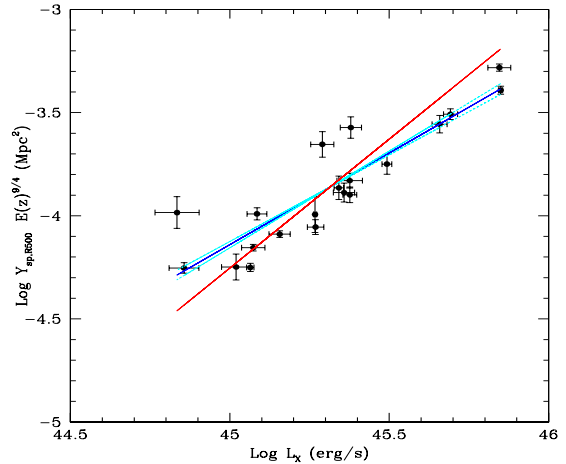}
\caption{We show here the best fit line (solid blue) together with the associated uncertainties in the slope and intercept (dashed cyan) and the theoretical expectation (solid red) for the $Y_{sph,R500}-L_X$ relation. 
The best-fit value of $X$ is $ 0.55 \pm 0.05 $ for the case of on isothermal $\beta$-model with a core radius of $ r_{c}=0.3 R_{500}$ and $\beta = 2/3$.}
\label{fig.4}
\end{figure}

The theoretical prediction for a constant value of $X$ for all clusters is shown in Fig.\ref{fig.4} together with the best-fit correlation of the data. We stress that the theoretical curve calculated under these assumptions is sensitively steeper than the power-law best-fit to the data. This is the result of having assumed a constant value of $X$ for all cluster X-ray luminosities in our model. A decreasing value of $X$ with the X-ray luminosity (or with the Compton parameter) as $X \sim L_X^{-\xi}$ can alleviate the problem providing a better agreement between the cluster formation scenario and the non-thermal phenomena in RH clusters. 
%

%\section{ Computing $X$ for each individual cluster}
%\newline
In order to analyze this point, we compute the value of $X$ for each individual cluster in our sample by using the relationship between the Compton parameter and the X-ray bolometric luminosity given above. Table \ref{tab.4} reports the values of $X$ calculated for the considered clusters assuming the previous $\beta$-model.
The error in $X$ is calculated from the error in the Luminosity and the Compton parameter. It is given by
\begin{equation}
\Delta X^2 = \bigg(\frac{\partial X}{\partial L_X} \Delta L_X\bigg)^2 + \bigg(\frac{\partial X}{\partial Y_{sph,R500}} \Delta Y_{sph,R500}\bigg)^2
\end{equation}
It is interesting that the analysis presented in this paper can provide a barometric test of the overall pressure structure in galaxy clusters that can be also useful for future studies.
\begin{table}
\centering
\caption{Clusters name and their corresponding calculated $X$ parameters}
\label{tab.4}
\begin{tabular}{c c }
\\
\hline\hline
Cluster 		& $X$ ($\beta$-model) \\
\hline
1ES0657 &0.16 \\
RXCJ2003 &-\\
A2744 & 7.17\\
A2163 & 0.61\\
A1300 & 2.03\\
A0665 & 0.33\\
A773 & 0.41\\
A2256 & 1.33\\
Coma & 0.322\\
A0520 & 0.50\\
A209 & -\\
A754 & 0.478\\
A401 & 0.349\\
A697 & 0.140\\
A781 & 3.07\\
A1995 & 0.61\\
A2034 & 0.52\\
A2218 & 3.79\\
A1689 & 0.42\\
MACSJ0717 & -\\
A1914 & 0.13\\
A2219 & 0.21\\
A2255 & 1.65\\
\hline
\end{tabular}
\end{table}

\begin{figure}
\centering
\includegraphics[width=90mm,height=110mm]{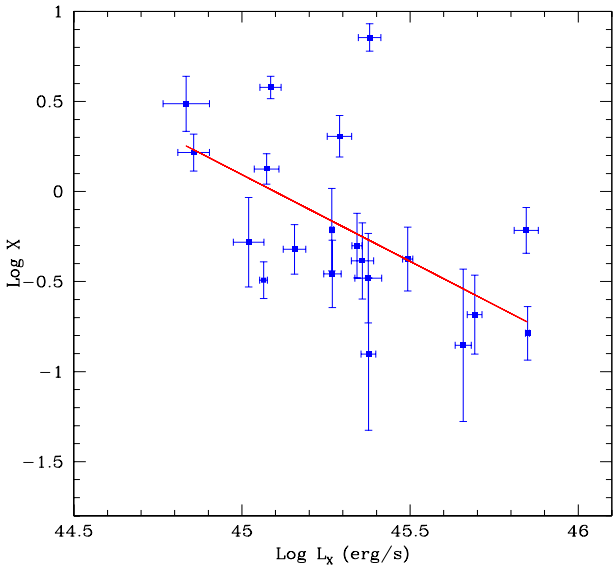}
\caption{The behavior of the pressure ratio $X$ as a function of the cluster X-ray bolometric luminosity for the cluster sample. The best fit curve $X  \sim L_X^{-0.96}$ is shown by the red solid line.}
\label{fig.5}
\end{figure}

Fig.\ref{fig.5} shows the correlation of the values of $X$ with both the Compton parameter and with the bolometric X-ray luminosity of each cluster. The data and our estimate for $X$ show that there is a clear decreasing trend of the pressure ratio $X$ with both the cluster X-ray luminosity and with the integrated Compton parameter indicating that low-$L_X$ (mass) cluster hosting RHs require a larger ratio of the non-thermal to thermal pressure ratio.
We fit the $X-L_X$ relation in Fig.\ref{fig.5} by assuming a power-law form
\begin{equation}
X = Q \cdot L_X^{-\xi}
\label{eq.x_vs_lx}
\end{equation}
and we obtain best fit values of $\xi = 0.96 \pm 0.16$ and $Log  Q = 43.49\pm 7.09$. The best fit curve with these parameters is also shown in Fig.\ref{fig.5}.
A $\chi_{red}^2= 1.14$ (with 17 d.o.f.)  is obtained in the case of $X \sim L_x^{-0.96}$ while a value $\chi_{red}^2= 1.56$ (with 18 d.o.f.) is obtained in the case $X =$const. This shows that the behaviour $X \sim L_x^{-0.96}$ is statistically significative: in fact, the probability of having a $\chi_{red}^2$  being larger than 1.14 (1.56) with 17 (18) d.o.f. is 0.307 (0.061).
The best-fit value of the exponent $\xi=0.96$ is different from $0$ at the 6 sigma confidence level.

\begin{figure}
\centering
\includegraphics[width=90mm,height=110mm]{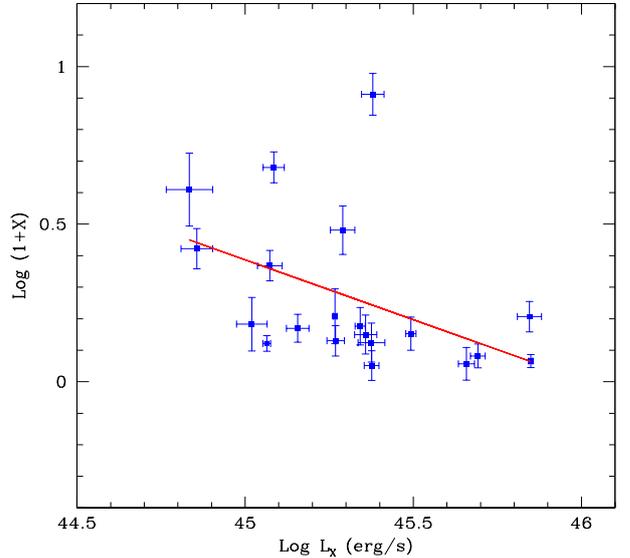}
\caption{The behavior of the total pressure ratio $1+X$ as a function of the cluster X-ray bolometric luminosity for the cluster sample. The best fit curve $1+X  \sim L_X^{-0.38}$ is shown by the red solid line. }
\label{fig.1pX_Lx}
\end{figure}

For the sake of completeness we also show in Fig.\ref{fig.1pX_Lx} the correlation between the total pressure ratio $1+X$ and $L_X$ that is fitted with a power-law of the form $(1+X) = Q' \cdot L_X^{-\xi'}$ with best-fit values $\xi' = 0.38 \pm 0.05$ and $Log   Q' = 17.50\pm 2.49$. Analogously, we find that a $\chi_{red}^2= 1.0$ (with 17 d.o.f.) is obtained in the case of $(1+X) \sim L_x^{-0.38}$ while a value $\chi_{red}^2= 1.33$ (with 18 d.o.f.) is obtained in the case $(1+X) =$const. Also in this case we find that the decrease of $1+X$ with the cluster X-ray luminosity is statistically significative: the probability of having a $\chi_{red}^2$  being larger than 1.00 (1.33) with 17 (18) d.o.f. is 0.454 (0.157).

\begin{figure}
\centering
\vbox{
\includegraphics[width=90mm,height=110mm]{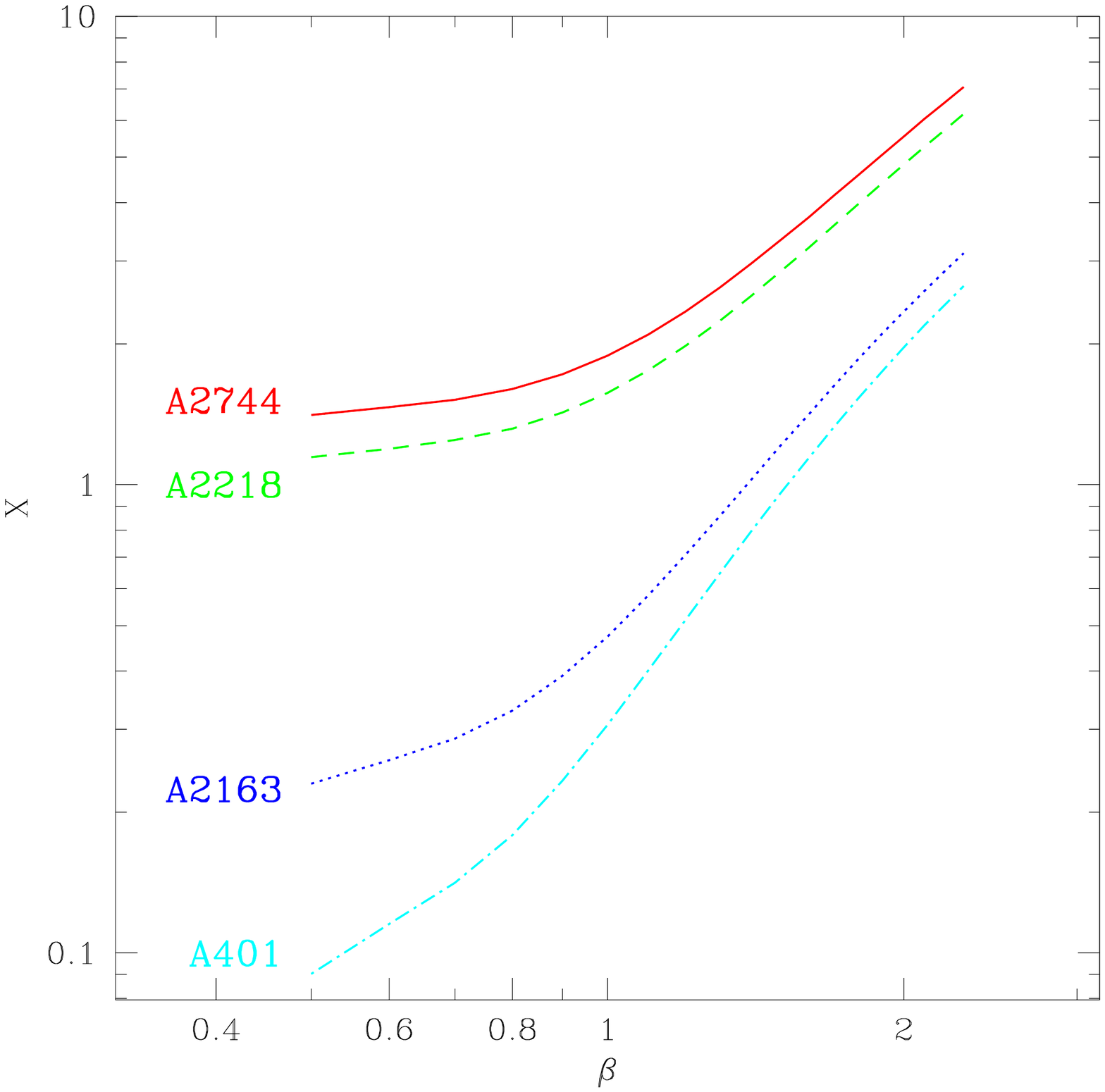}
\vskip -3.cm
\includegraphics[width=90mm,height=110mm]{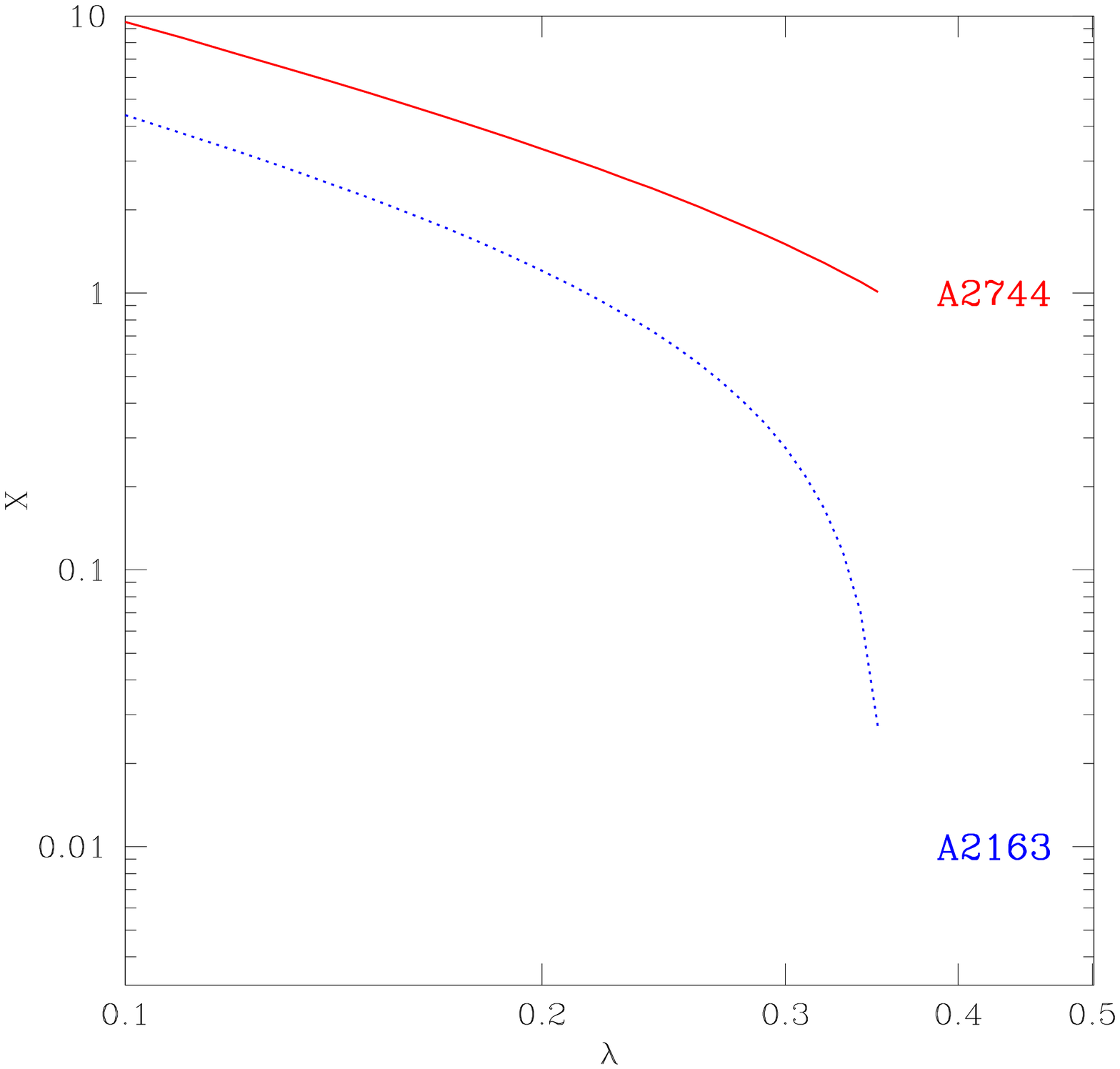}
}
\caption{The behavior of the pressure ratio $X$ as a function of the cluster $\beta$ parameter (upper panel) and as a function of the parameter $\lambda$ (lower panel) for some of the RH clusters in our list. }
\label{fig.7}
\end{figure}
\begin{figure}
\centering
\includegraphics[width=80mm,height=90mm]{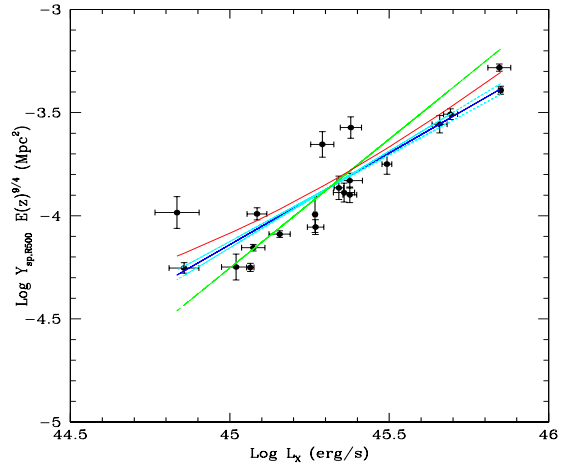}
\caption{We show here the best fit line to the data (solid blue) together with the associated uncertainties in the slope and intercept (dashed cyan) and the theoretical expectation (solid red curve). 
The relation $X \sim L_x^{-0.96}$ has been used in the theoretical prediction shown by the red solid curve.
The green dashed line is the theoretical prediction for $X=$ const.}
\label{fig.6}
\end{figure}

We then calculate our theoretical prediction for the $Y_{sph,R500} - L_X$ relation using the previous $X \propto L_X^{-\xi}$ relation and we find indeed a better agreement of the cluster formation model with the available data for our sample of RH clusters (see Fig.\ref{fig.6}).
This is confirmed by the reduced $\chi^2$ analysis. We have calculated the values of the $\chi^2_{red}$ in the two cases of  a constant value of $X$ and in the case in which we insert the relation $X \sim L_X^{-0.96}$, as it results from our analysis of the extended sample of clusters we consider after the Planck SZ catalog release. 
The $\chi_{red}^2$ for the case with a varying $X$, i.e. using the best fit $X \propto L_X^{-0.96}$,  is 0.96 (with 17 d.o.f.), while it is 0.86  (with 18 d.o.f.) in the case in which $X=$const.
This indicates that the fit to the data with a value of $X$ decreasing with the cluster X-ray luminosity is able to reproduce the correlation of the data better than in the case $X=$const., thus bringing consistency and robustness to our analysis and results.

Our result indicates that the existence of a non-thermal pressure in RH clusters with a ratio $X= P_{non-th}/P_{th}$ that decreases with cluster X-ray luminosity (or mass) is able to recover the consistency between the theoretical model for cluster formation and the presence of RHs in clusters.

\begin{figure}
\centering
\includegraphics[width=80mm,height=90mm]{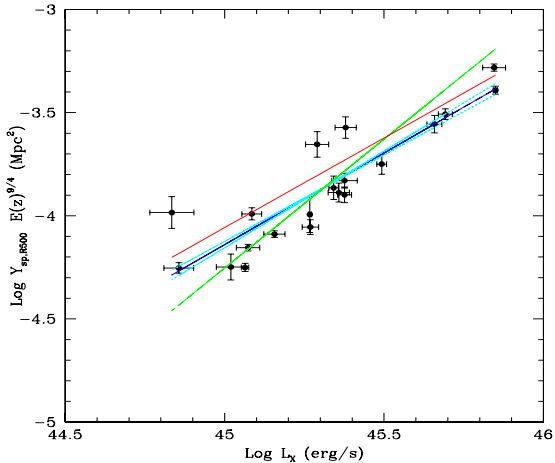}
\caption{We show here the best fit line to the data (solid blue) together with the associated uncertainties in the slope and intercept (dashed cyan) and the theoretical expectation (solid red curve). 
The relation $1+X \sim L_x^{-0.38}$ has been used in the theoretical prediction shown by the red solid curve.
The green line is the theoretical prediction for $1+X=$ const.}
\label{fig.YSZ_Lx_1px}
\end{figure}

Fig.\ref{fig.YSZ_Lx_1px} shows the theoretical prediction for the $Y_{sph,R500} - L_X$ relation using the best-fit correlation between the total pressure ratio and the bolometric X-ray luminosity, $1+X \propto L_X^{-0.38}$

Here the $\chi_{red}^2$ for the case in which we insert  the relation $1+X \propto L_x^{-0.38}$  is 0.97 (with 17 d.o.f.), while it is 0.86 (with 18 d.o.f.) in the case in which $1+X=$const. This again shows that the fit to the data with a value of $1+X$ decreasing with the cluster X-ray luminosity is  better that in the case $1+X=$const., and it is consistent with the best-fit analysis of the $(1+X) - L_X$ correlation.

\section{Discussion and conclusions}

We found evidence that the largest available sample of RH clusters with combined radio, X-ray and SZE data require a substantial non-thermal particle pressure to sustain their diffuse radio emission and to be consistent with the SZE and X-ray data.
This result has been derived mainly from the $Y_{sph,R500}-L_X$ relation for a sample of RH clusters selected from the Planck SZ effect survey.
This non-thermal particle (electron and positron) pressure affects in particular the value of the total Compton parameter $Y_{sph,R500}$ within $R_{500}$ indicating an integrated Compton parameter that is a factor $\sim 0.55\pm 0.05$ (on average) larger that the one induced by the thermal ICM alone.
The shape of the $Y_{sph,R500}-L_X$ does not depend on the assumptions on the cluster parameters and density profiles, while its normalization (and therefore the value of $X$) depend on the cluster parameters. Specifically, the value of $X$ decreases with increasing cluster core radius (or increasing value of $\lambda$) and increases with increasing value of the central particle density. Therefore, the normalization of the previous correlation, and consequently the best-fit value of $X$, are affected by the cluster structural parameters. Detailed studies of the values of $X$ derived from the previous correlation could be then used as barometric probes of the structure of cluster atmospheres.
However, one of the most important results we obtained in this work is that the simple description in which $X$ is constant for every cluster fails to reproduce the observed $Y_{sph,R500}-L_X$ relation, requiring that $X \sim L^{-0.96 \pm 0.16}_X$.  We hence found that the impact of the non-thermal particle pressure is larger (in a relative sense) in low-$L_X$ RH clusters than in high-$L_X$ RH clusters, requiring a luminosity evolution of the pressure ratio $X \sim L_X^{-\xi}$ with $\xi \approx 0.96 \pm 0.16$.
We note, in fact, that without this luminosity evolution the theoretical model for the $Y_{sphR500}-L_X$ correlation predicts a steeper relation compare to the best-fit one which is considerably flatter. 
A decreasing value of $X$ with the X-ray luminosity can therefore  provide a better agreement between the cluster formation scenario and the presence of non-thermal phenomena in RH clusters. 
This behavior can be attributed to the decreasing impact of the non-gravitational processes in clusters going from low to high values of $L_X$. 
This is consistent with a scenario in which relativistic electrons and protons are injected at an early cluster age by one or more cosmic ray sources and then diffuse and accumulate in the cluster atmosphere but are eventually diluted by the infalling (accreting) thermal plasma. This fact is also consistent with the outcomes of relativistic covariant kinetic theories of shock acceleration in galaxy clusters (see, e.g., Wolfe and Melia 2006, 2008) that predict that the major effect of shocks and mergers is to heat the ICM (rather than accelerating electrons at relativistic energies): in such a case the relative contribution of non-thermal particles to the total pressure in clusters should decrease with increasing cluster temperature, or X-ray luminosity since these two quantities are strongly correlated.  
Detailed models for the origin and distribution of the $P_{non-th}$ have to challenge the results present here and we will  discuss the relative phenomenology elsewhere (Colafrancesco et al. 2013, in preparation).

The positive values of $X$ found in our cluster analysis indicates the presence of a considerable non-thermal pressure provided by the non-thermal electrons (and positrons): the presence of non-thermal electrons (positrons) is the minimal particle energy density requirement because it has been derived from SZE measurements (i.e. by Compton scattering of CMB photons off high-energy electrons, and positrons). For a complete understanding of the overall cluster pressure structure one should also consider the additional contribution of non-thermal proton that is higher than the electron one since protons loose energy on a much longer time scale. Therefore, the derived values of $X$ should be considered as lower limits to the actual total non-thermal pressure and this will point to the presence of a relatively light non-thermal plasma in cluster atmospheres.  A full understanding of the proton energy density (pressure) in cluster atmospheres could be obtained by future gamma-ray observations (or limits) of these galaxy clusters with RHs because the gamma-ray emission could possibly be produced by $\pi^0 \to \gamma + \gamma$ decays where the neutral pions $\pi^0$ are the messengers of the presence of hadrons (protons) in cluster atmospheres (see, e.g., Colafrancesco \& Blasi 1998, Marchegiani et al. 2007, Colafrancesco \& Marchegiani 2008 and references therein). We will address the consequences of this issue on the high-E emission properties of RH clusters elsewhere (Colafrancesco et al. 2013, in preparation).

The results presented in this paper are quite independent on our assumptions of the cluster structural properties. Specifically, the slope of the $Y_{sph,R500} - L_X$ relation does not depend on the detailed shape of the cluster density profile, and hence the condition $X \sim L_X^{-0.96}$ seems quite robust. However, the absolute value of the pressure ratio $X$ for each cluster depends on the assumed density profile and on the simplifying assumption that the non-thermal electron distribution resembles the thermal ICM one. It might be considered, in general,  that the non-thermal and thermal particle density radial distributions are correlated as $n_{e,non-th} (r) \propto [n_{e,th}(r)]^{\alpha}$, and previous studies (see Colafrancesco and Marchegiani 2008) showed that the values of $\alpha$ do not strongly deviate from $1$, thus rendering our assumption reasonable and our result robust.

%A previous study of the $Y_{SZ}-L_X$ correlation has been provided by the Planck Collaboration (2011) and they found that the slope of this relation for the %whole set of SZ cluster in the Planck database is flatter than in the case of RH selected clusters considered here.

In conclusion, we have shown that the combination of observations on RH clusters at different wavelengths (radio, mm. and X-rays) is able to provide physical constraints on the non-thermal particle content of galaxy clusters. This is possible by combining the relevant parameters carrying information on the non-thermal (i.e. the total Compton parameter) and thermal (i.e. the X-ray bremsstrahlung luminosity) pressure components residing in the cluster atmosphere. The next generation radio (e.g. SKA and its precursors, like MeerKAT), mm. (e.g. Millimetron, and in general mm. experiment with spatially-resolved spectroscopic capabilities) and X-ray instruments will definitely shed light on the origin of radio halos in galaxy clusters and on their cosmological evolution.

\begin{acknowledgements}
S.C. acknowledges support by the South African Research Chairs Initiative of the Department of Science and Technology and National
Research Foundation and by the Square Kilometre Array (SKA). M.S. E., P.M. and N.M. acknowledge support from the DST/NRF SKA post-graduate bursary initiative.
\end{acknowledgements}

%%%%%%%%%

\end{document}